\documentclass{eptcs}

\usepackage[english]{babel}
\usepackage[utf8x]{inputenc}
\usepackage{graphicx}
\usepackage{booktabs}
\usepackage{url}
\usepackage{amsmath,amsfonts,amssymb}
\usepackage{dcolumn}
\usepackage{theorem}
\usepackage{microtype}
\usepackage{colonequals}
\usepackage[lined, boxed]{algorithm2e}
\usepackage{subfigure}
\usepackage{color}
\usepackage{todonotes}
\usepackage{xspace}
\newtheorem{definition}{Definition}
\newtheorem{example}{Example}
\newcommand{\naturals}{\ensuremath{\mathbb{N}}}
\newcommand{\reals}{\ensuremath{\mathbb{R}}}
\newcommand{\realsnn}{\reals_{\geq 0}}
\newcommand{\realsinf}{\realsnn^\infty}
\newcommand{\Prob}{\ensuremath{\mathbf{P}}}
\newcommand{\R}{\ensuremath{\mathbf{E}}}

\newcommand{\MAM}{\ensuremath{\mathcal{M}}\xspace}

\newcommand{\MS}{\ensuremath{\mathit{MS}}}
\newcommand{\PS}{\ensuremath{\mathit{PS}}}

\newcommand{\Distr}{\ensuremath{\text{Distr}}}
\newcommand{\RDistr}{\ensuremath{\text{RDistr}}}
\newcommand{\ub}{\ensuremath{\mathit{ub}}}
\newcommand{\lb}{\ensuremath{\mathit{lb}}}
\newcommand{\tb}{\ensuremath{\mathit{tb}}}
\newcommand{\pb}{\ensuremath{\mathit{pb}}}
\newcommand{\class}[2]{[#1]_{#2}}
\newcommand{\partition}{\ensuremath{\mathcal{P}}}
\newcommand{\pointdistr}[1]{\xi_{#1}}
\newcommand{\lambdamax}{\lambda_{\max}}
\newcommand{\erf}{\textsl{ER}_{\lambdamax,\tb}(\delta)}

\newcommand{\succset}{\mathrm{Succ}}

\newcommand{\dcup}{\mathbin{\dot\cup}}
\newcommand{\ie}{i.\,e.\xspace}
\newcommand{\eg}{e.\,g.\xspace}

\begin{document}
\title{
  MeGARA: \underline{Me}nu-based \underline{G}ame \underline{A}bstraction and\\ Abstraction \underline{R}efinement of Markov \underline{A}utomata
}

\author{Bettina Braitling\textsuperscript{1}\qquad Luis Mar\'{\i}a {Ferrer Fioriti}\textsuperscript{2}\qquad Hassan Hatefi\textsuperscript{2} \\ 
        Ralf Wimmer\textsuperscript{1}\qquad Bernd Becker\textsuperscript{1}\qquad Holger Hermanns\textsuperscript{2}
\institute{
  \textsuperscript{1}University of Freiburg, Germany \\
  \texttt{\{braitlin, wimmer, becker\}@} \\
    \texttt{informatik.uni-freiburg.de} \and
  \textsuperscript{2}Saarland University, Germany \\
  \texttt{\{ferrer, hhatefi\}@depend.cs.uni-saarland.de}  \\
  \texttt{hermanns@cs.uni-saarland.de}
}}
\def\titlerunning{MeGARA: Menu-based Game Abstraction and Abstraction Refinement for Markov Automata}
\def\authorrunning{B.\ Braitling, L.\ M.\ Ferrer Fioriti, H.\ Hatefi, R.\ Wimmer, B.\ Becker, H.\ Hermanns}

\maketitle
\renewcommand{\thefootnote}{\fnsymbol{footnote}} \footnotetext[1]{%
  This work was partly supported by the German Research Council (DFG)
  as part of the Transregional Collaborative Research Center
  ``Automatic Verification and Analysis of Complex Systems''
  (SFB/TR~14~AVACS), by the DFG/NWO Bilateral Research Programme
  ROCKS, and by the European Union Seventh Framework Programme under
  grant agreement no.\@ 295261 (MEALS) and 318490 (SENSATION).}
 \renewcommand{\thefootnote}{\arabic{footnote}}

% $Id: 00-abstract.tex  braitlin $
\begin{abstract}
  \noindent Markov automata combine continuous time, probabilistic
  transitions, and nondeterminism in a single model. They represent 
  an important and powerful way to model a wide range of complex real-life
  systems. However, such models tend to be large and difficult to handle,
  making abstraction and abstraction refinement necessary. In this paper
  we present an abstraction and abstraction refinement technique for
  Markov automata, based on the game-based and menu-based abstraction of
  probabilistic automata. First experiments show that a significant 
  reduction in size is possible using abstraction.
\end{abstract}

%%% Local Variables: 
%%% mode: latex
%%% TeX-master: "main"
%%% End: 
% $Id: 05-introductions.tex  braitlin $

\section{Introduction}
\label{sec:Introduction}

% introducing MA
Markov automata (MA) constitute a compositional behavioural model for
continuous-time stochastic and nondeterministic
systems~\cite{EisentrautHZ10:lics,EisentrautHZ10:concur,DengH13}.
MA are on one hand rooted in continuous-time Markov chains
(CTMCs) and on the other hand based on probabilistic automata
(PA)~\cite{Segala95}. MA have seen applications in diverse areas
where exponentially distributed delays are intertwined with
instantaneous random switching. The latter enables MA to capture the
complete semantics~\cite{EisentrautHKZ13} of generalised stochastic
Petri nets (GSPNs)~\cite{MarsanCB84} and of stochastic activity
networks (SANs)~\cite{MeyerMS85}. As MA extend Hermanns' interactive
Markov chains (IMCs)~\cite{Hermanns02}, they inherit IMC
application domains, ranging from GALS hardware
designs~\cite{CosteHLS09} and dynamic fault trees~\cite{BoudaliCS10}
to the standardised modelling language
AADL~\cite{BozzanoCKNNR11,HaverkortKRRS10}. Due to these attractive
semantic and compositionality features, there is a growing interest in
modelling and analysis techniques for MA.

% papers on MA
The semantics of MA including weak and strong (bi)simulation has been
studied
in~\cite{EisentrautHZ10:lics,EisentrautHZ10:concur,DengH13}. Markov
automata process algebra (MAPA)~\cite{TimmerKPS12} supports fully
compositional construction of MA equipped with some minimisation
techniques. Analysis algorithms for expected reachability time,
long-run average, and timed (interval) reachability have been studied
in~\cite{GuckHHKT13}.  It is also accompanied by a tool chain that
supports modelling and reduction of MA using SCOOP~\cite{TimmerKPS12}
and analysis of the aforementioned objectives using
IMCA~\cite{Guck12}.  Model checking of MA with respect to continuous
stochastic logic (CSL) has been presented in~\cite{HatefiH12}.

% time-bounded until, timed reachability
The core complexity of MA model checking lies in the model checking of
\emph{time-bounded until} formulae.  The property is reducible to
\emph{timed reachability} computation where the maximum and minimum
probability of reaching a set of target states within a time bound is
asked for. The current trend inspired by~\cite{Neuhausser10,ZhangN10}
is to split the time horizon into equally sized discretisation steps,
each small enough such that with high probability at most one Markov
transition occurs in any step. However, the practical efficiency and
accuracy of this approach turns out to be substantially inferior to
the one known for CTMCs, and this limits the applicability to real
industrial cases. It can only scale up to models with a few thousand
states, depending on the parameters of the model and the time bound
under consideration. This paper proposes an abstraction refinement
technique to address the scalability problem of model checking
time-bounded until for MA.

% menu-based game abstraction
Abstraction refinement methods have gained popularity as an effective
technique to tackle scalability problems (\eg state space
explosion) in probabilistic and non-probabilistic settings. Although
abstraction refinement techniques have not been employed for MA yet,
there are a number of related works on PA, which allow to estimate
lower and/or upper bounds of reachability
probabilities. \emph{PA-based abstraction}~\cite{DArgenioJJL02}
abstracts the concrete model into a PA, which provides an upper
bound for maximal and a lower bound for minimal reachability of the
concrete model. \emph{Game-based abstraction}~\cite{KattenbeltKNP10}
on the other hand enables to compute both lower and upper bound on
reachability, \ie the reachability probability in PA is guaranteed to
lie in a probability interval resulting from the analysis of the
abstract model, which is represented by a game. This further has been
proven to be the best transformer~\cite{WachterZ10}. In this work we
nevertheless employ \emph{menu-based abstraction}~\cite{Wachter2011},
which can be exponentially smaller in the size of transitions and also
easier to implement. Moreover it provides lower and upper bounds on
reachability like game-based abstraction.

% overview of the work
In this paper we introduce a menu-based game abstraction refinement
approach which generalises Wachter's method \cite{Wachter2011} to MA and combines it with
Kattenbelt's method~\cite{KattenbeltKNP10}. As mentioned before, the essential part of CSL
model checking reduces to timed reachability computation. We
accordingly focus on this class of properties. We exploit
\emph{scheduler-based refinement} which splits an abstract block by
comparing the decisions made by the lower and upper bound schedulers.
Furthermore, we equip the refinement procedure with a pseudo-metric 
that measures how close a scheduler is to the optimal one. It turns
out that the latter enhances the splitting procedure to be coarser. 
We start the computation with a relatively low precision and increase 
it repeatedly, thus speeding up the refinement procedure in the 
beginning while ensuring a high quality of the final abstraction.
Our experiments show promising results, especially we can report 
on a significant compaction of the state space.

\noindent\textit{Organisation of the paper.} At first we give a brief
introduction into the foundations of MA and stochastic games in
Section~\ref{sec:Foundations}. Afterwards we will present our approach
for the menu-based game abstraction of MA, its analysis and subsequent
refinement in Section~\ref{sec:AbstractionAnalysisRefinement}.
Experimental results will be shown in Section~\ref{sec:Experiments}.
Section~\ref{sec:Conclusion} concludes the paper and gives an outlook
to future work.

% $Id: 02-foundations.tex  braitlin $

\section{Foundations}
\label{sec:Foundations}

In this section we will take a brief look at the basics of MA 
and continuous-time stochastic games.

\subsection{Markov Automata}
\label{sec:MarkovAutomata}

We denote the real numbers by $\reals$, the non-negative real numbers
by $\realsnn$, and by $\realsinf$ the set $\realsnn\cup\{\infty\}$.
For a finite or countable set $S$ let $\Distr(S)$ denote the set of
probability distributions on $S$, \ie\ of all functions 
$\mu:S\to[0,1]$ with $\sum_{s\in S}\mu(s)=1$. A rate distribution on
$S$ is a function $\rho:S\to\realsnn$. The set of all rate
distributions on $S$ is denoted by $\RDistr(S)$.

\begin{definition}[Markov automaton]
  \label{def:ma}
  A \emph{Markov automaton} (MA) $\MAM = (S,s_0,A,\Prob,R)$ consists of
  a finite set $S$ of states with $s_0\in S$ being the initial state,
  a finite set $A$ of actions, a \emph{probabilistic transition relation}
  $\Prob \subseteq S \times A \times \Distr(S)$, and a \emph{Markov transition relation}
  $R: S\to\RDistr(S)$.
\end{definition}

The rate $R(s)(s')$ is the parameter of an exponential distribution
governing the time at which the transition from state $s$ to state $s'$ becomes
enabled. The probability that this happens within time $t$ is given by
$1-e^{-R(s)(s')\cdot t}$.

We make the usual assumption that we have a \emph{closed system},
\ie all relevant aspects have already been integrated into
the model such that no further interaction with other components
occurs. Then nothing prevents probabilistic transitions from being
executed immediately. This is called the \emph{maximal progress
assumption}~\cite{DengH13}. Since the probability that a Markov
transition becomes enabled immediately is zero, we may assume that a
state has either probabilistic or Markov transitions. We denote the
set of states with Markov transitions as $\MS$. $\PS$ is the set of
states with probabilistic transitions. It holds that 
$\MS \cap \PS =\emptyset$.

If there is more than one Markov transition leaving $s\in\MS$
a \emph{race condition} occurs~\cite{DengH13}:
The first transition that becomes enabled is taken.
We define the exit rate $\R(s) = \sum_{s'\in S} R(s)(s')$
of state $s\in\MS$.

Starting in the initial state $s_0$, a run of the system is generated
as follows: If the current state $s\in\MS$ is Markovian, the sojourn time
is determined according to the continuous probability distribution
$(1-e^{-\R(s)t})$. At this point in time a transition to $s'\in S$ occurs
with probability $\frac{R(s)(s')}{\R(s)}$. Taking this together, the
probability that a transition from $s\in\MS$ to $s'$ occurs within time
$t\geq 0$ is 
\[
  \mu(s)(s',t) = (1-e^{-\R(s)t})\frac{R(s)(s')}{\R(s)}\ .
\]

In a probabilistic state $s \in \PS$ first a transition
$(s,\alpha,\mu)\in\Prob$ is chosen nondeterministically. Then the
probability to go from $s$ to successor state $s'\in S$ is given by
$\mu(s')$. The sojourn time in probabilistic states is 0.

The nondeterminism between the probabilistic transitions in state $s$
is resolved through a \emph{scheduler}.  The most general scheduler
class maps the complete history up to the current probabilistic state
to the set of transitions enabled in that state. Considering the
general scheduler class is extremely excessive for most objectives
like time-bounded reachability, for which a simpler class, namely
\emph{total-time positional deterministic schedulers}
suffice~\cite{Neuhausser10}. Schedulers of this class resolve
nondeterminism by picking an action of the current state, which is
probabilistic, based on the total time that has elapsed. Formally, it
is a function $\sigma:\PS \times \mathbb{R}_{\ge0} \rightarrow
A\times\Distr(S)$ with $\sigma(s,t)=(\alpha,\mu)$ only if
$(s,\alpha,\mu)\in\Prob$.

% generic scheduler
% \begin{definition}[Generic Scheduler]\label{gms:def}
%   A generic scheduler over MA $\MAM = ( S, Act, \longrightarrow,
%   \dashrightarrow, \nu )$ is a function, $A: \fpaths \times Act_{\bot}
%   \rightarrowtail [0,1]$, where $A(\pi,.) \in
%   \dists(Act_{\bot}(\lasts(\pi))), \pi \in \fpaths$. Scheduler $A$ is
%   measurable if $\forall \alpha \in Act_{\bot}, A(.,\alpha): \fpaths
%   \rightarrowtail [0,1]$ is measurable.
% \end{definition}
% For a finite path $\pi$, a scheduler specifies how to resolve
% nondeterminism in the last state of $\pi$. If $\lasts(\pi)$ is a
% Markov state, there will be only one enabled action, namely $\bot$,
% which will be chosen by the scheduler with probability 1. Otherwise,
% it gives a distribution over the set of enabled actions of
% $\lasts(\pi)$. Measurability of scheduler $A$ means the preimage of
% each Borel measuable set in $A$ is a measuable set of paths,
% \ie $\forall \alpha \in Act_{\bot}, \forall B \in
% \mathfrak{B}([0,1]), \; \left \{ \pi | A(\pi, \alpha) \in B \right \}
% \in \mathfrak{F}_{Paths^*}$, where $\mathfrak{B}([0,1])$ is Borel
% $\sigma$-algebra over interval $[0,1]$. We use term \GMS to refer to
% the set of all generic measurable schedulers.

% time-bounded reachability 
Time-bounded reachability in MA quantifies the minimum and the maximum
probability to reach a set of target states within a given time
interval. A fixed point characterisation is proposed
in~\cite{HatefiH12} to compute this objective. The characterisation
is, however, in general not algorithmically
tractable~\cite{BaierHHK03}. To circumvent this problem, the fixed
point characterisation is approximated by a discretisation
approach~\cite{HatefiH12}. Intuitively, the time horizon is divided
into equally sized sub-intervals, each one of length
$\delta>0$. Discretisation step $\delta$ is presumed to be small
enough such that, with high probability, at most one Markov transition
fires within time $\delta$.  This assumption discretises an MA by
summarising its behaviour at equidistant time points. Time-bounded
reachability is then computed on the discretised model, together with
a stable error bound. The whole machinery is here generalised to
stochastic games and the algorithm is later employed to establish a
lower and an upper bound for both minimal and maximal time-bounded reachability 
probabilities in MA.

The MA we consider are non-Zeno, \ie they do not have any
end components consisting only of probabilistic
states. Otherwise it would be possible to have an infinite amount of
transitions taking place in a finite amount of time.

For more on MA in general we recommend~\cite{DengH13}.

\subsection{Stochastic Games}
\label{sec:StochasticGames}
Stochastic games are generalisations of MA. They also
combine continuous time with nondeterminism and probabilities.

A stochastic game consists of one or several players who can choose
between one or several actions. In turn, these actions may influence
the behaviour of the other players. Each action consists of a
real-valued or infinite rate $\lambda\in\realsinf$ and a probability distribution. For our
work we need the definition of two-player games:

\begin{definition} [Stochastic game] 
\label{def:StochasticGame}
  A \emph{stochastic continuous-time two-player game} is a tuple 
  $\mathcal{G} = (V,(V_1,V_2),$ $v_0, A, T)$ such that
  $V = V_1\dcup V_2$ is a set of states,
  $v_0\in V$ is the initial state,
  $A$ is a finite set of actions and
  $T \subseteq V \times A \times\realsinf\times \Distr(V)$ 
  is a probabilistic transition relation with continuous time.
\end{definition}

$V_1$ and $V_2$ are the states of player~1 and player~2, respectively. We
define two functions $\theta_p:T\to\Distr(V)$ and $\theta_r : T \to
\realsinf$. $\theta_p$ is a projection on the probability
distribution of a transition, $\theta_r$ is a projection on the rate.
If the current state is $v\in V_1$, then it is player~1's turn to choose the next transition, 
otherwise player~2's. The current player chooses a transition $(v,\alpha,\lambda,\mu)\in T$ 
for leaving state $v$. $\theta_r(v,\alpha,\lambda,\mu) = \lambda \in \realsinf$
determines how long this action takes, whereas
$\theta_p(v,\alpha,\lambda,\mu) = \mu \in \Distr(V)$ gives us the
distribution which leads to a successor state. A typical goal of such
games is, \eg, that player~1 wants to reach a goal state within a
given time bound and player~2 tries to prevent this.

In the following we denote states of player~1 and player~2 as $V_1$-states 
and $V_2$-states.

The nondeterminism which may occur at a certain player state is
resolved by a scheduler, which is in this case called a \emph{strategy}. Each player follows his
own strategy in order to accomplish its goal. As for MA, total-time
positional deterministic strategies are sufficient since we are
concentrating on time-bounded reachability. A strategy for player
$x\in \{1,2\}$ is therefore defined as a function $\sigma_x: V_x
\times \realsnn\to A\times\realsinf\times\Distr(V)$, with
$\sigma_x(v,t)=(\alpha,\lambda,\mu)$ only if
$(v,\alpha,\lambda,\mu)\in T$. Depending on their strategies, players
may co-operate or compete with each other.

With the strategies of both players in place, the nondeterminism 
within a stochastic game is resolved, the result being a deterministic 
MA. Stochastic games with strategies therefore have the same semantics 
as MA, especially the discretisation of continuous-time stochastic games 
works in a similar way.

For more on strategies and on stochastic games in general we refer 
to~\cite{Shapley1953}.

%%% Local Variables: 
%%% mode: latex
%%% TeX-master: "main"
%%% End: 

% $Id: 03-foundations.tex  braitlin $

\section{Abstraction, Analysis and Refinement}
\label{sec:AbstractionAnalysisRefinement}

\emph{Abstraction} in general is based on a partition $\partition =
\{B_1, B_2, \ldots, B_n\}$ of the state space. The original or
\emph{concrete} states are lumped together into abstract states,
defined by the blocks $B_i\in\partition$. 

For PA, both game- and menu-based abstraction use these blocks $B_i$ as player~1 states
(\ie $V_1=\partition$). In game-based abstraction~\cite{KattenbeltKNP10} for PA 
player~2 states in $V_2$ represent sets of concrete
states that have the same branching structure.
In menu-based abstraction~\cite{Wachter2011} the states of
player~2 represent the set of enabled actions within a block
$B_i$. Abstraction refinement for both approaches is based on values
and schedulers which are computed for certain properties.

MA are an extension of PA, they additionally contain Markov
transitions. In our work we aspire to transfer the results
of~\cite{KattenbeltKNP10} and~\cite{Wachter2011} from PA to MA. Menu-based
abstraction~\cite{Wachter2011} is usually more compact than
game-based abstraction~\cite{KattenbeltKNP10}, since in general
there are more different states within a block than different enabled
actions. However, the game-based abstraction is more suitable for
Markovian states as Markov transitions are not labelled with
actions. Therefore we decided to combine both techniques, which is
described in the following section.

For the remainder of the paper we define $A(B) = \{ \alpha\in A\,\vert\,
  \exists s\in B\ \exists\mu\in\Distr(S): (s,\alpha,\mu)\in \Prob\}$ as the 
set of actions which are enabled within a set of states $B \subseteq S$.

\subsection{Menu-based Game Abstraction for MA}
\label{sec:Abstraction}

As in the case of~\cite{KattenbeltKNP10, Wachter2011}, our
\emph{menu-based game abstraction} is based on a partition
$\partition$. Each block of $\partition$
either contains probabilistic or Markovian states, not both. It holds
that $B_i \cap B_j = \emptyset$ for all $i,j \in \{1 \ldots, n\}$.

The probabilistic blocks of partition $\partition$ constitute the $V_1$-states,
whereas $V_2$-states either represent the enabled actions within a probabilistic
block, a Markovian block of $\partition$, or the concrete states within a
Markovian block. Thus, the original nondeterminism of the MA is represented
in the $V_1$-states, the nondeterminism artificially introduced by the abstraction
is present in $V_2$-states.

The transitions of the original MA $\MAM$ have to be
\emph{lifted} to sets of states as follows:
\begin{definition} [Lifted (rate) distribution]
\label{def:Lifting}
  Let $\mu\in\Distr(S)$
  be a probability distribution over $S$ and $\partition$ a partition of
  $S$. The \emph{lifted distribution}
  $\overline{\mu}\in\Distr(\partition)$ is given by $\overline{\mu}(B) = \sum_{s\in B}\mu(s)$
  for $B\in\partition$. Accordingly for a rate distribution 
  $\rho\in\RDistr(S)$ we define $\overline{\rho}\in\RDistr(\partition)$ by
  $\overline{\rho}(B) = \sum_{s\in B}\rho(s)$ for all $B\in\partition$.
\end{definition}

\begin{figure}[tb]
\vspace{-0.25cm}
   \centering
   \subfigure[]{
      \label{fig:menu-probabilistic_state_2_1}
      \includegraphics[height=3.75cm]{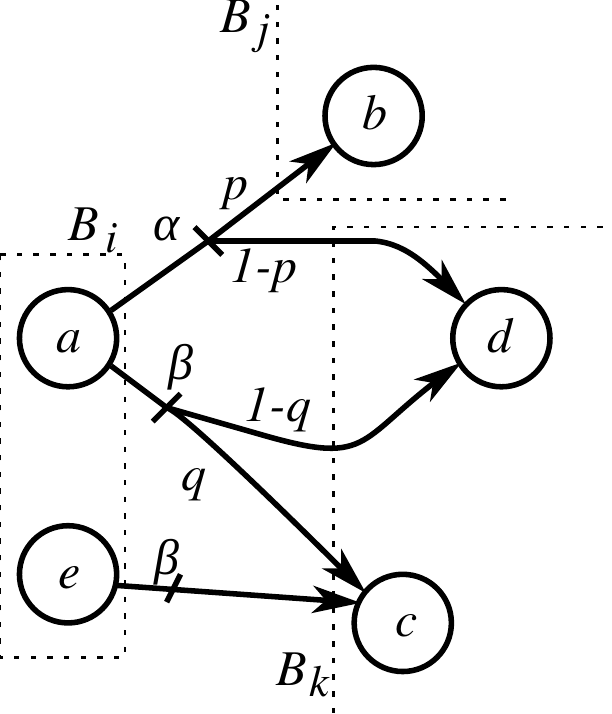}
    }
\hspace*{0.5cm}    
   \subfigure[]{
      \label{fig:menu-probabilistic_state_2_2}
      \includegraphics[height=3.0cm]{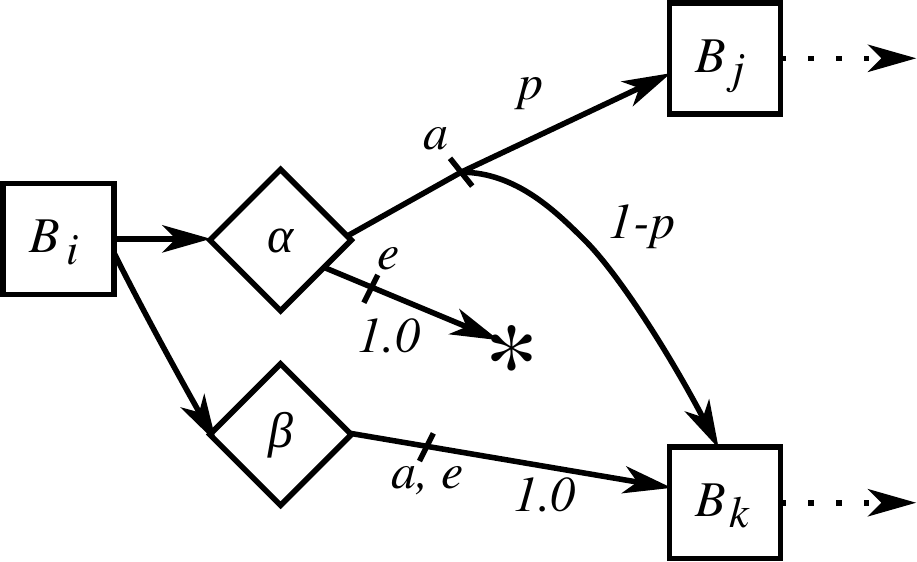}
    }
    \caption{Example for the menu-based abstraction of probabilistic states.}
   \label{fig:menu-probabilistic_state}
\end{figure}

If several probabilistic distributions for an action $\alpha$ within a partition block $B_i$
turn out to be the same after lifting, they are unified. Additionally, if action
$\alpha\in A(B_i)$ is not enabled in a state $s \in B_i$, then a new probabilistic
distribution is added with $\pointdistr{*}(*) = 1.0$, `$*$' being
a newly added bottom state. This can be interpreted as the lifting of
nonexistent distributions.
Example~\ref{ex:ProbabilisticState} and Figure~\ref{fig:menu-probabilistic_state} illustrate the abstraction process
for probabilistic states, which is a direct transfer from Wachter's
menu-based abstraction~\cite{Wachter2011}.

\begin{example}
\label{ex:ProbabilisticState}
Figure~\ref{fig:menu-probabilistic_state_2_1} shows a part of an MA $\MAM$ 
and a partition $\partition$. The probabilistic states `$a$' and `$e$' 
of $\MAM$ are contained in the same block $B_i$. In the abstraction, 
which is shown in Figure~\ref{fig:menu-probabilistic_state_2_2}, $B_i$ 
becomes a $V_1$-state---indicated as a square---, whereas the actions 
$\alpha, \beta \in A(B_i)$ become $V_2$-states---indicated as diamonds. 
Since $\alpha$ is not enabled in the concrete state `$e$', a `$*$'-transition 
is added to the abstract $\alpha$-state.
\end{example}

As can be seen, the original nondeterminism is resolved by the choice of player~1 
between the different enabled actions, whereas an introduced nondeterminism is 
present at player~2. To make the later abstraction refinement easier 
(s.\ Section~\ref{sec:Refinement}) we also retain a mapping between the 
abstract distributions and the corresponding concrete states, as indicated 
in Figure~\ref{fig:menu-probabilistic_state_2_2}.

For Markovian states, the menu-based approach from~\cite{Wachter2011} cannot be 
used, since Markov transitions do not have actions which can be used as $V_2$-states. 
This is indicated by using the $\bot$-symbol. Our approach for Markovian states is therefore 
more similar to Kattenbelt's game-based abstraction~\cite{KattenbeltKNP10}: A Markovian block $B_i$ becomes a $V_2$-state, succeeded by $V_2$-states representing the concrete states within $B_i$ which have the same lifted Markov transitions
according to Definition~\ref{def:Lifting}. In the following, we denote $V_2$-states representing a Markovian block (concrete states) as \emph{abstract} (\emph{concrete}) Markovian $V_2$-states.

\begin{figure}[tb]
   \centering
   \subfigure[]{
      \label{fig:markovian_state_2_1}
      \includegraphics[height=3.75cm]{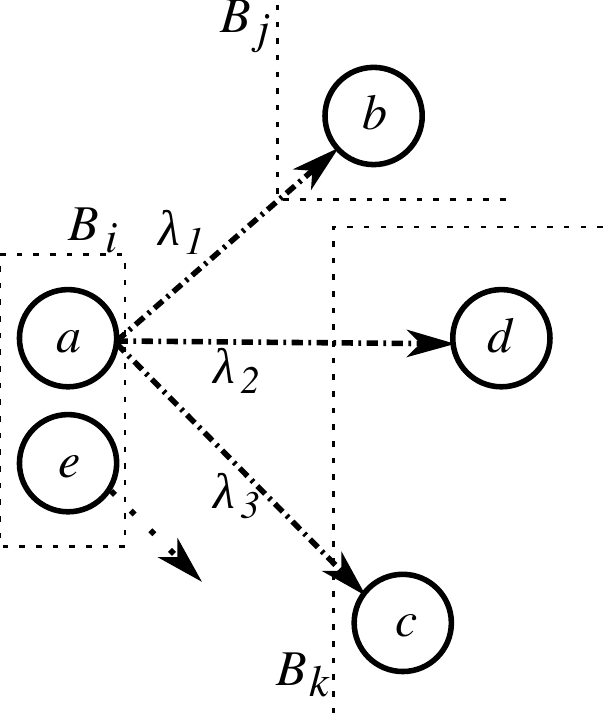}
   }
   \hspace*{0.5cm}   
   \subfigure[]{
      \label{fig:markovian_state_2_2}
      \includegraphics[height=3.0cm]{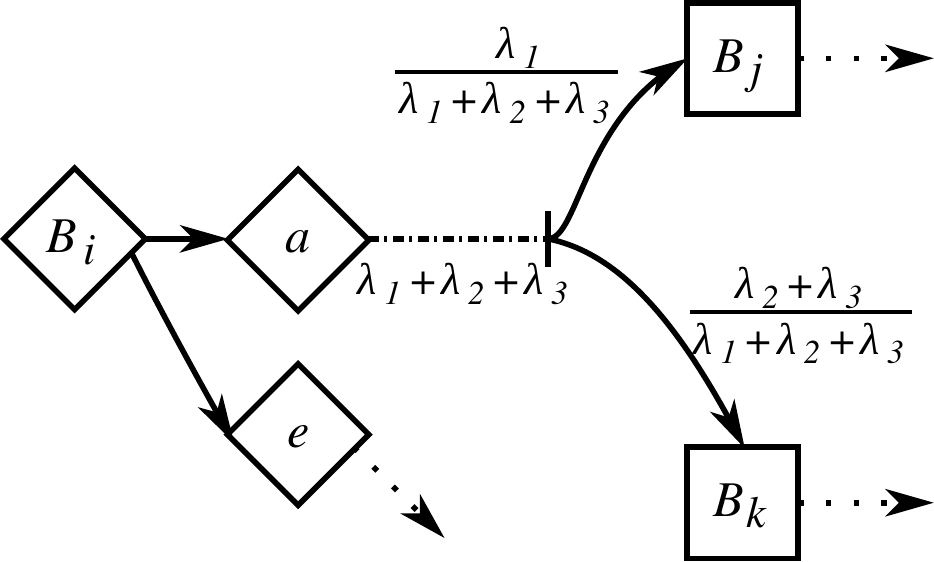}
   }
   \caption{Example for the game-based abstraction of Markovian states.}
   \label{fig:markovian_state}
\end{figure}

Example~\ref{ex:MarkovianState} and Figure~\ref{fig:markovian_state} demonstrate the abstraction 
of Markovian states.
\begin{example}
\label{ex:MarkovianState}
Figure~\ref{fig:markovian_state_2_1} shows a part of an MA $\MAM$. The Markovian 
states `$a$' and `$e$' of $\MAM$ are contained in the same block $B_i$. 
$\lambda_1$, $\lambda_2$ and $\lambda_3$ denote the rates of the rate distribution of 
`$a$'.

The block $B_i$ becomes an abstract Markovian $V_2$-state in the abstraction as 
shown in Figure~\ref{fig:markovian_state_2_2}. The other $V_2$-states, however, correspond directly 
to the concrete states. Since the lifted rate distribution of `$a$' is different from the 
one of `$e$'---not shown here---, the concrete states stay separate, otherwise they would 
be unified.
\end{example}

It has to be noted, that at abstract Markovian $V_2$-states only introduced nondeterminism occurs.
In the concrete system there is no nondeterminism here, we have the race condition instead.

All transitions leading from a $V_1$-state $v_1$ to a $V_2$-state $v_2$ are considered to be 
\emph{immediate}, \ie they do not require any time. This is symbolised by giving them 
the rate $R(v_1)(v_2) = \infty$. The same holds for transitions leading from abstract Markovian to concrete Markovian $V_2$-states and for probabilistic transitions from a probabilistic $V_2$-state to a successor state. The nondeterministic transitions from a 
$V_1$- or from an abstract Markovian $V_2$- to a $V_2$-state $v_2$ are associated with the unique probability distribution 
$\pointdistr{v_2}$ with $\pointdistr{v_2}(v_2) = 1.0$. For a clearer representation we 
omitted point-distributions $\pointdistr{v_2}$ and rates $\infty$ in the preceding 
and in the following figures and examples.

We additionally define $\class{s}{\partition}, s\in S$ as the (unique) block $B\in\partition$ with $s\in B$.

After these preliminaries we can formally define our menu-based game abstraction of MA.

\begin{definition} [Menu-based game abstraction]
  \label{def:Abstraction}
  Given an MA $\MAM =  (S,s_0,A,\Prob,R)$ and partition $\partition = \{B_1, \dots B_n\}$ of $S$. We construct 
  the menu-based game abstraction $\mathcal{G}_\MAM^\partition = (V, (V_1, V_2), v_0, \overline{A}, \overline{T})$ with:
  \begin{itemize}
  \item $V = V_1 \dcup V_2$, 
  \item $V_1 = \{v\in\partition\,|\,v\subseteq\PS\} \dcup \{ * \}$,
  \item $V_2 = \bigl\{ (v_1, \alpha)\in \partition\times A\,\big\vert\, v_1 \subseteq \PS\land \alpha \in  A(v_1) \bigr\} 
     \dcup \bigl\{v\in\partition\,|\,v\subseteq\MS\bigr\}$ \\
     \hspace*{2em}${} \dcup 
     \bigl\{(v_1,\overline{\rho})\in\partition\times\RDistr(\partition)\,\big\vert\, 
     v_1\subseteq\MS\land\exists s\in v_1: R(s)=\rho\bigr\}$,
  \item $v_0 = \class{s_0}{\partition}$,
  \item $\overline{A} = A \dcup\{\bot\}$, and
  \item $\overline{T} \subseteq V \times \overline{A} \times \realsinf \times \Distr(V)$ is
    given by $\overline{T} = \overline{T_{\PS}}\dcup \overline{T_{\MS}}$ with
    \begin{align*}
      \overline{T_{\PS}} & = \bigl\{ (\class{s}{\partition},\bot,\infty,\pointdistr{(\class{s}{\partition},\alpha)})\,\big\vert\, s\in\PS\land\alpha\in A(s)\bigr\} \\
      &\qquad\cup \bigl\{ \bigl( (\class{s}{\partition},\alpha), \alpha, \infty, \overline{\mu}\bigr)\,\big\vert\, s\in\PS\land (s,\alpha,\mu)\in P\bigr\} \\
      &\qquad\cup \bigl\{ \bigl( (\class{s}{\partition},\alpha), \bot, \infty, \pointdistr{\ast}\bigr)\,\big\vert\, s\in\PS\land \alpha\not\in A(s)\bigr\} \\
      \overline{T_{\MS}} &= \bigl\{ (\class{s}{\partition},\bot,\infty,\pointdistr{(\class{s}{\partition},\overline{\rho})})\,\big\vert\, s\in\MS\land \rho = R(s)\bigr\} \\
      &\qquad\cup \bigl\{ \bigl( (\class{s}{\partition},\overline{\rho}),\bot, \R(s), \frac{\overline{\rho}}{\R(s)}\bigr)\,\big\vert\, s\in\MS\land \rho=R(s)\bigr\}.
    \end{align*}
 \end{itemize}
\end{definition}

The probability distributions $\mu \in \Distr(V)$ and rate distributions 
$\rho \in \RDistr(\partition)$ are as stated previously. For the remainder 
of this paper we will not refer to the transition relation $\overline{T}$, 
but directly to the probabilistic distributions and rate distributions.

\begin{figure}[tb]
   \centering
   \subfigure[]{
      \label{fig:menu-queueing_system_2}
      \includegraphics[height=3.75cm]{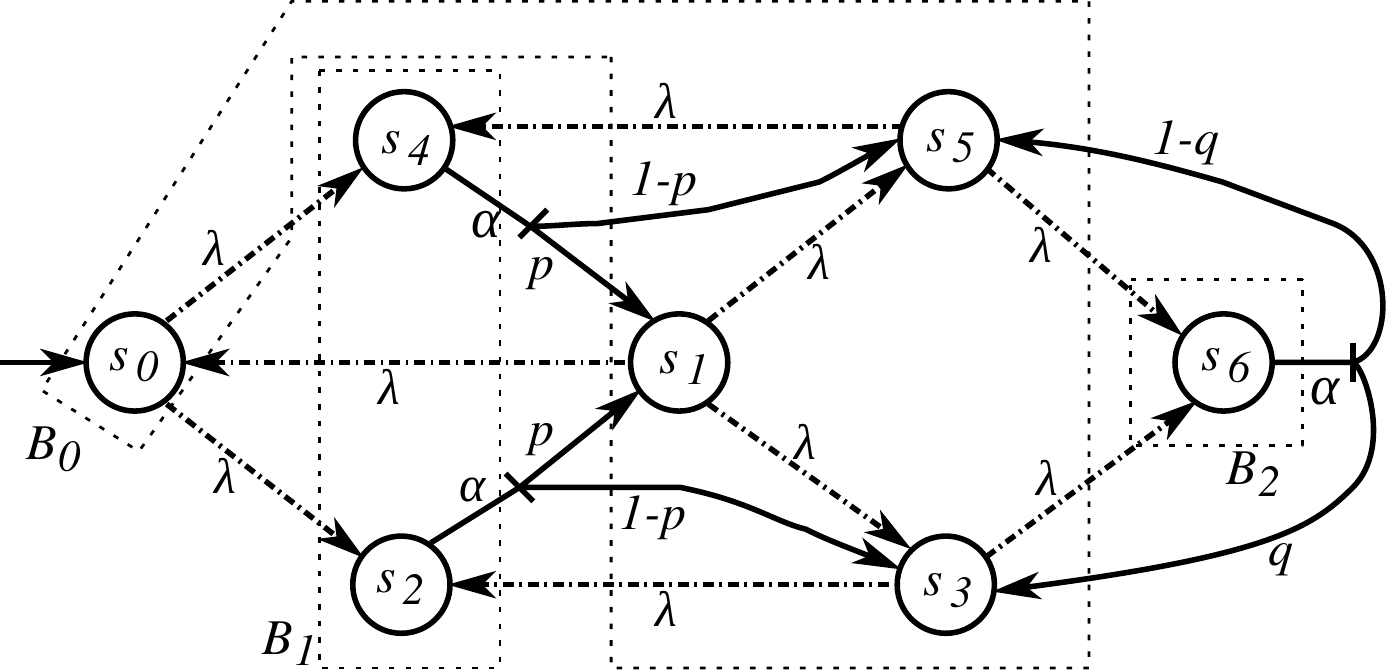}
   }
\hspace*{\fill}
   \subfigure[]{
      \label{fig:menu-qs_abstracted_2}
      \includegraphics[height=4.25cm]{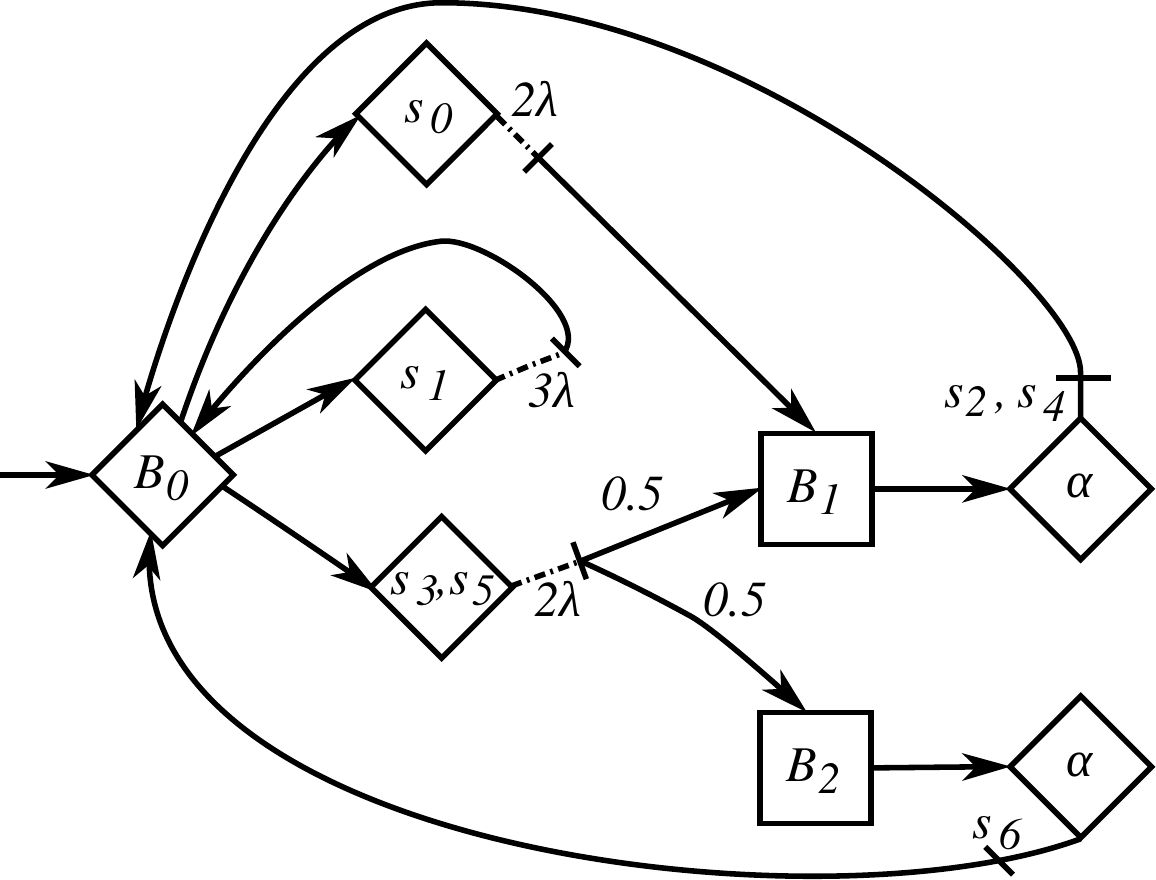}
   }
   \caption{An example for the menu-based game abstraction of an MA $\MAM$.}
   \label{fig:abstraction-qs}
\end{figure}

Figure~\ref{fig:abstraction-qs} and Example~\ref{ex:abstraction_qs} illustrate the abstraction process.
\begin{example}
\label{ex:abstraction_qs}
Figure~\ref{fig:menu-queueing_system_2} shows an MA $\MAM$ and a partition 
$\partition = \{B_0, B_1, B_2\}$. $B_0$ contains all Markovian states of 
$\MAM$ and $B_1$ all probabilistic states, except the goal state 
$s_6$, which is contained in the separate block $B_2$. The corresponding 
menu-based game abstraction $\mathcal{G}_\MAM^\partition$ is pictured in 
Figure~\ref{fig:menu-qs_abstracted_2}.

As can be seen, $B_0$ becomes an abstract Markovian $V_2$-state, whereas the blocks $B_1$ and $B_2$ build the $V_1$-states of $\mathcal{G}_\MAM^\partition$. The abstract Markovian state $B_0$ leads to a set of $V_2$-states, which either correspond directly to concrete states (in the case of `$s_0$' and `$s_1$') or to a set of concrete states (in the case of `$s_3, s_5$'). The abstract Markovian $V_2$-state $B_0$ is also the initial state of $\mathcal{G}_\MAM^\partition$, since the block contains the concrete initial state $s_0$.

The only enabled action within block $B_1$ is $\alpha$, the same holds for $B_2$. The corresponding abstract states lead each to a $V_2$-state labelled with $\alpha$. Although $B_1$ contains two probabilistic states, $s_2$ and $s_4$, only one distribution goes out from the respective $\alpha$-state, since the distributions after lifting are identical.
\end{example}

\subsection{Analysis of the Abstraction}
\label{sec:Analysis}

As mentioned before, there are two kinds of nondeterminism present in the
abstraction: the original, concrete nondeterminism and the introduced abstract
one. As in~\cite{KattenbeltKNP10, Wachter2011} the nondeterministic choices are
resolved by two separate schedulers: the \emph{concrete scheduler} 
$\sigma_c:V_1\times\realsnn\to\overline{T}$ of player~1
and the \emph{abstract scheduler} $\sigma_a:V_2\times\realsnn\to\overline{T}$ of player~2. 
These schedulers are total-time positional deterministic strategies of stochastic
games (see Section~\ref{sec:StochasticGames}).

For Markovian blocks only the abstract scheduler $\sigma_a$ exists, since only the introduced nondeterminism is present there. 
This can be seen in Figure~\ref{fig:markovian_state}: A (nondeterministic) choice occurs at the abstract Markovian $V_2$-state 
only, whereas there is no choice for the concrete Markovian $V_2$-states.

While the concrete scheduler $\sigma_c$ always behaves according to the property under consideration, \eg in case of maximal bounded reachability $\sigma_c$ tries to maximise the result, the abstract scheduler $\sigma_a$ can either co-operate or compete with $\sigma_c$, \ie it can try to maximise or minimise the probability. This leads to the existence of an upper and a lower bound for every property. The value of the original system lies within these bounds. We omit the proof of this for now, however it is similar to the proof of the correctness of the menu-based or game-based abstraction of PA in~\cite{KattenbeltKNP10} or in~\cite{Wachter2011}.

If the bounds are too far apart, the abstraction is too coarse and has to be refined (s.\ Section~\ref{sec:Refinement}).

As already mentioned, we are currently concentrating on time-bounded reachability, but we are going to consider a wider set of properties in the future.

\subsubsection{Time-bounded Reachability}
\label{sec:TimeBoundedReachability}

If we want to analyse a property within the abstraction $\mathcal{G}_{\MAM}^{\partition}$, we have to compute lower and upper bounds for this property. For example for the maximum probability $p_{\max}^{\mathcal{G}_{\MAM}^\partition}$ to reach a set of goal states $\overline{G}$ within time bound $\tb$, starting at a state $v \in V$, we get:
\begin{align*}
  p_{\max,\lb}^{\mathcal{G}_{\MAM}^\partition}(v,\Diamond^{\le\tb} \overline{G}) &= \underset{\sigma_c}{\sup}\,\underset{\sigma_a}{\vphantom{p}\inf}\ \mathit{Pr}_{v,\sigma_c,\sigma_a}(\Diamond^{\le\tb} \overline{G}), \\
  p_{\max,\ub}^{\mathcal{G}_{\MAM}^\partition}(v,\Diamond^{\le\tb} \overline{G}) &= \underset{\sigma_c}{\sup}\,\underset{\sigma_a}{\sup}\ \mathit{Pr}_{v,\sigma_c,\sigma_a}(\Diamond^{\le\tb} \overline{G})\ ,
\end{align*}
where $\mathit{Pr_{v,\sigma_c,\sigma_a}}$ is the probability measure
induced on the abstraction by the state $v$ and the two schedulers
$\sigma_c$ and $\sigma_a$. $\Diamond^{\le\tb}$ is the
"Finally"-operator as known from linear temporal logic (LTL), bounded
to time interval $[0,\tb]$. For the remainder of this paper we will
concentrate on the maximum probability
$p_{\max,\pb}^{\mathcal{G}_{\MAM}^\partition}(v,\Diamond^{\le\tb}
\overline{G})$ and abbreviate it as $p_{\max,\pb}(v,\tb)$, for
$\pb\in\{\lb,\ub\}$. The computation of the minimum probability is
analogous.

For the remainder of the paper we define $\succset(v) = \bigl\{v' \in V \,\vert\, (v,\alpha,\lambda,\overline{\mu}) \in \overline{T} \land \overline{\mu}(v') \neq 0\bigr\}$ as the set of successor states of a state $v \in V$.

% Fixed point characterisation
The maximum (minimum) reachability probabilities can be computed
similarly to MA~\cite{HatefiH12} by using a fixed point
characterisation. Formally, $p_{\max,\pb}(v,\tb)$ is the least fixed
point of higher-order operator
$\Omega_{\max,\pb}:(V\times\realsnn\rightarrow[0,1])\rightarrow(V\times\realsnn\rightarrow[0,1])$:
\begin{align}
\intertext{For $v\in V_2,\, v = (v_1,\overline{\rho}) \in\partition\times\RDistr(\partition)$:}
\label{eq:FixedpointMarkovV2}
\Omega_{\max,\pb}(F)(v,\tb) &= \begin{cases}
\displaystyle\int_{0}^{\tb} \R(v)e^{-\R(v)t} \sum\limits_{v'\in V_1}\overline{\rho}(v') F(v',\tb-t)\,\mathrm{d}t, & \text{if } v\notin \overline{G},\\
1, & \text{if } v\in \overline{G}. \\
\end{cases} \\
\intertext{For $v \in V_2,\, v = (v_1,\alpha)\in \partition\times A$:}
\label{eq:FixedpointProbabilisticV2}
\Omega_{\max,\pb}(F)(v,\tb) & = \begin{cases}
  1, & \text{if } v\in \overline{G},\\
  \underset{s \in v_1,}{\max} \, \sum\limits_{v' \in V_1} \overline{\mu}(v')F(v',\tb), & \text{if } v \notin \overline{G},\, \pb = \ub, \\
  \underset{s \in v_1}{\min} \, \sum\limits_{v' \in V_1} \overline{\mu}(v')F(v',\tb), & \text{if } v \notin \overline{G},\, \pb = \lb. \\
  \end{cases} \\
\intertext{for $v \in V_2,\, v \subseteq \MS$:}
\label{eq:FixedpointMarkovV1}
\Omega_{\max,\pb}(F)(v,\tb) &= \begin{cases}
  \underset{v' \in \succset(v)}{\max}\, F(v',\tb), & \text{if } \pb = \ub, \\
  \underset{v' \in \succset(v)}{\min}\, F(v',\tb), & \text{if } \pb = \lb. \\
  \end{cases} \\  
\intertext{For $v \in V_1,\ v \subseteq \PS,\ (v,\alpha)\in V_2$:}
\label{eq:FixedpointProbabilisticV1}
\Omega_{\max,\pb}(F)(v,\tb) &= \underset{\alpha \in A(v)}{\max}\, F\bigl((v,\alpha),\tb\bigr)\ .
\end{align}
As can be seen, the recursive computation of the probability ends when
a goal state $g \in \overline{G}$ is reached. Therefore it is
sound to make goal states absorbing prior to the computation. The
fact that concrete Markovian $V_2$-states do not have a
nondeterministic choice is reflected in
Equation~\eqref{eq:FixedpointMarkovV2} of the definition of
$\Omega_{\max,\pb}$. In this case it does not matter whether $\lb$ or
$\ub$ is computed.  The fixed point characterisation implicitly
computes an optimal concrete and an optimal abstract scheduler.  The
schedulers are total-time positional deterministic as follows from
Equations~\eqref{eq:FixedpointProbabilisticV2},
\eqref{eq:FixedpointMarkovV1} and
\eqref{eq:FixedpointProbabilisticV1}. In each of the equations the
optimal choice, which depends solely upon current state $v$ and time
instant $\tb$, is deterministically---not randomly---picked.

The time bound $\tb$ only affects Markov transitions. Nevertheless, 
the resulting equation system is usually not algorithmically tractable, 
as is the case for MA. As for MA, we therefore approximate the result by 
using \emph{discretisation}~\cite{HatefiH12}, which we will discuss 
in the next section.

\subsubsection{Discretisation}
\label{sec:Discretisation}
The interval $[0,\tb]$ is split into $n\in \naturals$
discretisation steps of size $\delta>0$, \ie $\tb=n\cdot\delta$. The
discretisation constant $\delta$ has to be small enough such that, with
high probability, at most one Markov transition occurs within time
$\delta$. The probability distributions in this case have to be adjusted. 
For a concrete Markovian $V_2$-state $v$ and a state $v' \in V$ we get:
\[
\overline{\mu}_{\delta}(v') = \begin{cases}
  (1 - e^{-\R(v)\delta})\, \overline{\rho}(v') + e^{-\R(v)\delta}, & \text{if  $v'$ is the (unique) predecessor of $v$,}\\
  (1 - e^{-\R(v)\delta})\, \overline{\rho}(v'), & \text{otherwise.} \\
  \end{cases}
\]

In the first case of $\overline{\mu}_{\delta}$ a new transition from the $V_2$-state 
$v$ to its preceding abstract Markovian $V_2$-state $v'$ is added, if no such transition already exists.

% An additional error is added through the discretisation, however we will skip its analysis at this point. The error is in $O((\lambda_{\max}\delta)^2)$, similar as the error for the discretisation of MA~\cite{GuckHHKT13}, with $\lambda_{\max}$ being the biggest real-valued rate in the abstraction.
% finding a safe step-size  
An additional error is added through the discretisation, however we
will skip its analysis at this point. The error is at most
$\erf=1-e^{\lambda_{\max}\tb}(1+\lambda_{\max}\delta)^n$, similar to
the error for the discretisation of MA~\cite{GuckHHKT13}, with
$\lambdamax$ being the biggest real-valued rate in the abstraction.
Given a predefined accuracy level $\epsilon$, a proper step size
$\delta$ can be computed such that $\erf\le\epsilon$. A simple
solution is to use the linear approximation
$n\frac{(\lambdamax\delta)^2}{2}$, which is a safe upper bound of the
error function, \ie $\erf\le
n\frac{(\lambdamax\delta)^2}{2}$. However, this is not a good
approximation when the value of the error function is not close to
zero.  In such a case, it is worthwhile to use Newton's step method to find
a proper step size $\delta$ based on precision $\epsilon$. This leads to a smaller number of iterations
without violating the accuracy level.

% Given precision $\epsilon$, a proper step size $\delta$ can be computed by doing Newton's step.
Subsequently, the discrete-time menu-based game abstraction
$\mathcal{G}_{\MAM,\delta}^\partition$ is induced. Given a time bound
$\tb$ and a set of target states $G$, we can compute a lower and an
upper bound of the maximum probability to reach the states in $G$
within time bound $\tb$ for the discrete game, denoted by
$\tilde{p}_{\max,\lb}$ and $\tilde{p}_{\max,\ub}$ respectively, using
a value iteration algorithm.  At each discrete step, the algorithm
computes the optimal choice of each player on the discrete game,
thereby implicitly providing hop-counting positional deterministic
concrete and abstract schedulers, \ie deterministic schedulers
deciding based on the current state and the length of the path visited
so far.  The schedulers establish an $\epsilon$-optimal approximation
for reachability of the original game, \ie $\forall v\in
V.~\tilde{p}_{\max,\pb}(v,\tb)\le
p_{\max,\pb}(v,\tb)\le\tilde{p}_{\max,\pb}(v,\tb)+\epsilon$, with
$\pb\in\{\lb,\ub\}$. For lack of space we have to leave the exact
algorithm out, a similar one for MA can be found in~\cite{HatefiH12}.

\subsection{Abstraction Refinement}
\label{sec:Refinement}
We are currently using a \emph{scheduler-based} refinement technique, similar to the strategy-based refinement of~\cite{KattenbeltKNP10} and the refinement technique from~\cite{Wachter2011}, which uses pivot blocks.
The key idea in these refinement techniques is the fact that a difference in the lower and upper bound probabilities in the abstraction requires that the abstract schedulers $\sigma_a^\lb$ and $\sigma_a^\ub$ differ in at least one abstract state.
Clearly fixing a scheduler for player 2 transforms the stochastic Markov game into an MA, and it has a unique maximum probability.
Thus any refinement strategy based on the previous observation can be reduced to (1) finding a set of abstract states where $\sigma_a^\lb$ and $\sigma_a^\ub$ disagree, and
(2) splitting these abstract states using some well-defined procedure.
For the first part we take all probabilistic or abstract Markovian $V_2$-states $v$ that have different strategies for the lower and the upper bound, and are reachable from the initial state from one of the two composed schedulers, and also have probabilities which differ by more than $\epsilon$. We denote this state set as $D_{\sigma_a}$.

Splitting the abstract states is more involved.
Given a state $v\in D_{\sigma_a}$, its preceding $V_1$- or $V_2$-state $B$ and a transition $t=(v,\alpha,\lambda, \overline{\mu}) \in \overline{T}$, the set of concrete states $B_t$ is defined as $\{s \in B\, \vert\, \mu = \overline{\mu} \land \lambda=\R(s)\}$. Notice that $\partition_{B, \alpha} = \{B_t \,\vert\, t = (v, \alpha, \lambda, \overline{\mu}) \in \overline{T} \}$ is a partition of $B$. One possibility is to split $B$ using $\partition_{B, \alpha}$, but this can introduce a lot of new abstract states that are irrelevant for the abstraction.
The approach of \cite{KattenbeltKNP10, Wachter2011}  is to replace $B$ by the sets $B_{v, \sigma_a^\lb}$, $B_{v, \sigma_a^\ub}$, and $B\setminus (B_{v, \sigma_a^\lb} \cup B_{v, \sigma_a^\ub})$. Although this removes the choices that caused the divergence in the scheduler, it certainly does not remove all similar divergences that can arise in the refined abstraction.
That is the case when $B\setminus (B_{v, \sigma_a^\lb} \cup B_{v, \sigma_a^\ub})$ contains choices with probabilities close to the lower and upper bound probabilities $p_{\max,\lb}$ and $p_{\max,\ub}$.
Our approach consists in splitting $B$ using a bounded pseudo-metric $m$ over distributions.
A pseudo-metric $m : \Distr(V) \times \Distr(V) \to [0,1]$ satisfies $m(\mu, \mu) = 0$, $m(\mu_1, \mu_2) = m(\mu_2, \mu_1)$ and $m(\mu_1, \mu_2) \leq m(\mu_1, \mu') + m(\mu', \mu_2)$.
So, if $v$ is a $V_2$-state such that $\mu^\lb = \sigma_a^\lb(v) \neq \sigma_a^\ub(v) = \mu^\ub$ and $m(\mu^\lb, \mu^\ub) = d$, then we split $B$ into $B_{v, \mu^\lb, \frac{d}{2}}$, $B_{v, \mu^\ub, \frac{d}{2}}$ and $B \setminus (B_{v, \mu^\lb, \frac{d}{2}} \cup B_{v, \mu^\ub, \frac{d}{2}})$, where $B_{v, \mu, d} = \bigcup \{B_{v, \lambda, \mu'} \,\vert\, m(\mu, \mu') \leq d\}$.
The pseudo-metric we adopt is $m(\mu, \mu') = \sum \vert \mu(v) - \mu'(v) \vert (\tilde{p}_{\max,\ub}(v,\tb)- \tilde{p}_{\max,\lb}(v,\tb))$. More precise and sophisticated metrics can be used, \eg the Wasserstein metric that has the property that all bisimilar distributions have distance 0 \cite{DesharnaisJGP02} (in our metric distance zero implies bisimilarity).
% \bcb{One reviewer requested more details about the use of the pseudo-metric.
% 
% @Luis: Could you provide that?}
% 
\begin{algorithm}[tb]
\label{alg:AbsRefLoop}
%\SetAlogLined
\KwIn{An MA $\MAM$, a set of goal states $G$, a time bound $\tb$, a desired precision $\epsilon$}
\KwOut{A menu-based game abstraction $\mathcal{G}_\MAM^\partition$ such that $p_{\max,\ub}(v_0,\tb)-p_{\max,\lb}(v_0,\tb)\le\epsilon$}
\BlankLine
$\partition \leftarrow \{B_1, \hdots, B_n\}$ such that $B_1 = G$\;
$\hat{\epsilon} \leftarrow 1$\;
$\textit{done} \leftarrow \textit{false}$\;
\While{$!\textit{done}$}{
	build $\mathcal{G}_\MAM^\partition$ from $\partition$\;
        find step size $\delta$ such that $\erf\le\hat{\epsilon}$ using Newton's step method\;
        discretise $\mathcal{G}_\MAM^\partition$ into $\mathcal{G}_{\MAM,\delta}^\partition$\;
	compute $\tilde{p}_{\max,\lb}(\cdot,\tb)$, $\sigma^{\lb}_a$, $\sigma^{\lb}_c$ for $\mathcal{G}_{\MAM,\delta}^\partition$ using value iteration\;
	compute $\tilde{p}_{\max,\ub}(\cdot,\tb)$, $\sigma^{\ub}_a$, $\sigma^{\ub}_c$ for $\mathcal{G}_{\MAM,\delta}^\partition$ using value iteration\;
	\lIf{$\tilde{p}_{\max,\ub}(v_0,\tb)-\tilde{p}_{\max,\lb}(v_0,\tb) + \hat{\epsilon}\leq \epsilon$}{
		$\textit{done} \leftarrow \textit{true}$
	}
	\lElseIf{$\tilde{p}_{\max,\ub}(v_0,\tb)-\tilde{p}_{\max,\lb}(v_0,\tb) \leq  \epsilon$}{
		$\hat{\epsilon} \leftarrow \max(\hat{\epsilon}/2, \hat{\epsilon}-\epsilon)$
	}
	\lElse{
		$\textit{Refine}(\mathcal{G}_\MAM^\partition, \partition, \sigma^{\lb}_a, \sigma^{\lb}_c, \sigma^{\ub}_a, \sigma^{\ub}_c)$
	}
}
\caption{Refinement Algorithm}
\end{algorithm}

Another novel approach used in our refinement algorithm is changing the precision when calculating the upper and lower bounds in the abstraction.
The number of iterations required is a function of $\epsilon$, the precision needed in the discretisation. The smaller $\epsilon$, the smaller step size, thus the larger number of iterations is required. Each iteration amounts to calculate an bounded reachability over an PA or a stochastic game.
If the maximum probability in the discretised concrete model is $p$, then the real probability is guaranteed to be in $[p, p+\epsilon]$. It is in turn over-approximated in the abstraction by $[\tilde{p}_{\max,\lb},\tilde{p}_{\max,\ub}+\epsilon]$.
If the abstraction is too coarse and consequently needs to be refined, then $\tilde{p}_{\max,\lb}$ and $\tilde{p}_{\max,\ub}$ can be obtained using the maximum $\hat{\epsilon}>\epsilon$ that triggers the refinement loop.
Algorithm \ref{alg:AbsRefLoop} shows the implementation of our abstraction refinement loop.

As mentioned earlier, the value of the concrete MA $\MAM$ for a certain property lies between the lower bound $p_{\max,\lb}$ and upper bound $p_{\max,\ub}$ of the menu-based game abstraction $\mathcal{G}_\MAM^\partition$. To evaluate the quality of the abstraction, the game needs to be discretised. Therefore, an appropriate step size $\delta$, which respects accuracy level $\hat{\epsilon}$, is computed using Newton's step method. Afterwards, the game is discretised into $\mathcal{G}_{\MAM,\delta}^\partition$, which is then analysed with respect to the given target set $G$ and the time bound $\tb$. We utilise the difference $d = \tilde{p}_{\max,\ub} - \tilde{p}_{\max,\lb}$ as a criterion of the current abstraction quality and compare it with the desired precision $\epsilon$. If $d+\hat{\epsilon}$ exceeds $\epsilon$, we refine our abstraction, \ie we refine the partition $\partition$. The result of this refinement step is a new menu-based game abstraction $\mathcal{G}_\MAM^\mathcal{P'}$ for which in turn new 
upper 
and lower bounds $\tilde{p}_{\max,\ub}'$ and $\tilde{p}_{\max,\lb}'$ can be computed. As soon as $d+\hat{\epsilon}$ is below $\epsilon$, we can stop the refinement process. The smaller we choose $\epsilon$, the more precise is the final result.

\subsubsection{Zenoness}
\label{sec:Zenoness}
Even if there is no probabilistic end component present in the original MA $\MAM$, it may happen that Zenoness is introduced into $\mathcal{G}_{\MAM}^\partition$, \eg through a non-cyclic chain of probabilistic states which are partitioned into the same block. Although probabilistic end components represent unrealistic behaviour -- it is possible to execute an infinite number of transitions in a finite amount of time -- in the case of time-bounded reachability it is not necessary to treat them separately. They will be dissolved automatically during refinement.

If we compute the lower bound $p_{\max,\lb}$ of $\mathcal{G}_{\MAM}^\partition$, the probability of a probabilistic end component without a goal state is $0$, because the goal state cannot be reached. Since goal states are made absorbing for the computation of time-bounded reachability, we do not have to consider the case that a goal state is contained within a probabilistic end component.

If we compute the upper bound $p_{\max,\ub}$, the probability of the end component is also $0$ (a goal state cannot be reached). In order to maximise its value, scheduler $\sigma_a$ will not select transitions leading into the end component and Zeno behaviour will be avoided.

Probabilistic end components are therefore only a problem when computing the lower bound, which will lead to $p_{\max,\lb} = 0$. This is the extreme value for $p_{\max,\lb}$ and unless the upper bound $p_{\max,\ub}$ is very low, \ie $p_{\max,\ub} \leq \epsilon$, the refinement loop will be triggered.

%%% Local Variables: 
%%% mode: latex
%%% TeX-master: "main"
%%% End:

% $Id: 04-experiments.tex  braitlin $

\section{Experimental Results}
\label{sec:Experiments}
We implemented in C++ a prototype tool based on our menu-based game abstraction, together with an analysing and refinement framework. For refinement we use the techniques we described in Section~\ref{sec:Refinement}. As mentioned earlier, we are currently considering bounded reachability objectives only, using discretisation (s. Section~\ref{sec:Discretisation}).

For our experiments we used the following case studies: 

 \begin{table}[tb]
  \caption{Maximum time-bounded reachability}
  \label{tab:ResultTable}
  \centering\scriptsize
   \begin{tabular}{|l|l||rrr|rrrrrr|}
   \hline
	& & \multicolumn{3}{c|}{Concrete Model} & \multicolumn{6}{c|}{Abstraction} \\
   Name & $\tb$ & \#{}states & $p$ & time & \#{}states & $\lb$ & $\ub$ & \#{}iter. & ref. time & val. time \\
   \hline
   PrG-2-active      & 5.0 & 2508 & 1.000 & 6:34 & 13 & 1.000 & 1.000 & 5 & 0:00 & 0:57 \\
   PrG-2-active-conf & 5.0 & 1535 & 1.000 & 2:38 & 6 & 1.000 & 1.000 & 5 & 0:00 & 0:19 \\
   PrG-2-empty       & 5.0 & 2508 & 0.993 & 5:37 & 394 & 0.992 & 0.993 & 19 & 0:01 & 7:59 \\
   PrG-2-empty-conf  & 5.0 & 1669 & 0.993 & 2:11 & 288 & 0.993 & 0.993 & 25 & 0:01 & 3:27 \\
%    PrG-3-active      & 5.0 & 10852 & 1.000 & 121:38 & 13 & 1.000 & 1.000 & 5 & 0:00 & 1:07 \\ 
%    PrG-3-active-conf & 5.0 & 6693 & 1.000 & 35:46 & 6 & 1.000 & 1.000 & 5 & 0:00 & 0:21 \\ 
   PrG-4-active      & 5.0 & 31832 &  \multicolumn{2}{c|}{(TO)} & 9 & 1.000 & 1.000 & 5 & 0:00 & 0:58 \\
   PrG-4-active-conf & 5.0 & 19604 & 1.000 & 113:31 & 6 & 1.000 & 1.000 & 5 & 0:00 & 0:21 \\ 
   \hline
   PoS-2-3      & 1.0 & 1497 & 0.557 & 0:02 & 508 & 0.555 & 0.557 & 17 & 0:00 & 0:04 \\
   PoS-2-3-conf & 1.0 & 990 & 0.557 & 0:01 & 443 & 0.557 & 0.557 & 19 & 0:00 & 0:02 \\
   PoS-2-4      & 1.0 & 4811 & 0.557 & 0:11 & 1117 & 0.557 & 0.557 & 17 & 0:00 & 0:13 \\
   PoS-2-4-conf & 1.0 & 3047 & 0.557 & 0:07 & 891 & 0.556 & 0.558 & 15 & 0:01 & 0:07 \\
   PoS-3-3      & 1.0 & 14322 & 0.291 & 1:38 & 5969 & 0.291 & 0.291 & 57 & 0:02 & 2:03 \\ 
   PoS-3-3-conf & 1.0 & 9522 & 0.291 & 1:15 & 5082 & 0.291 & 0.292 & 81 & 0:04 & 2:23 \\
%    PoS-4-2      & 1.0 & 6667 & 0.118 & 0:32 & 1729 & 0.117 & 0.119 & 25 & 0:00 & 0:20 \\ 
%    PoS-4-2-conf & 1.0 & 4745 & 0.118 & 0:16 & 3227 & 0.116 & 0.118 & 24 & 0:00 & 1:10 \\ 
   \hline
   GFS-20        & 0.5 & 7176 & 1.000 & 20:15 & 3 & 1.000 & 1.000 & 4 & 0:01 & 0:45 \\
   GFS-20-hw-dis & 0.5 & 7176 & 0.950 & 28:31 & 3164 & 0.950 & 0.950 & 26 & 0:34 & 36:16 \\
   GFS-30        & 0.5 & 16156 & 1.000 & 187:50 & 3 & 1.000 & 1.000 & 4 & 0:00 & 3:36 \\
   GFS-30-hw-dis & 0.5 & 16156 & 0.950 & 162:01 & 2412 & 0.950 & 0.950 & 23 & 9:28 & 120:09 \\
   GFS-40        & 0.5 & 28736 & \multicolumn{2}{c|}{(TO)} & 3 & 1.000 & 1.000 & 4 & 0:01 & 22:54 \\
%    GFS-40-hw-dis & 0.5 & 28736 & \multicolumn{2}{c|}{(TO)} & \multicolumn{6}{c|}{(TO)} \\
   GFS-50        & 0.5 & 44916 & \multicolumn{2}{c|}{(TO)} & 3 & 1.000 & 1.000 & 4 & 0:06 & 50:09 \\ 
  \hline
   \end{tabular}
   \end{table}

\noindent(1)~The \emph{Processor Grid} (PrG)~\cite{GuckHHKT13,TimmerPS13} consists of a $2\times 2$-grid of processors, each being capable of processing up to $K$ tasks in parallel. We consider two scenarios defined by two different set of goal states: Either the states in which the task queue of the first processor is empty or the states in which the first processor is active. Besides of the original model we also consider variants which were already compacted through the confluence reduction of~\cite{TimmerPS13}. The model instances are denoted as ``PrG-$K$-(active$\vert$empty)(-conf)''.

\noindent(2)~The \emph{Polling System} (PoS)~\cite{GuckHHKT13,TimmerPS13} consists of two stations and one server. Requests are stored within two queues until they are delivered by the server to their respective station. We vary the queue size $Q$ and the number of different request types $J$. As for PrG, we consider the original model as well as variants with confluence reduction. The goal states $G$ are defined as the states in which both station queues are full. The model instances are denoted as ``PoS-$Q$-$J$(-conf)''.

\noindent(3)~The \emph{Google File System}~\cite{GhemawatGL03,Guck12} (GFS) splits
files into chunks of equal size, maintained by several chunk servers.
If a user wants to access a chunk, it asks a master server that stores
the addresses of all chunks. Afterwards the user has direct read/write access on the chunk. For our experiments we fixed the number of chunks a server may store ($C_{\textsl{s}}=5000$), as well as the total number of chunks ($C_{\textsl{t}}=100000$), and we vary the number of chunk servers $N$. The set of goal states $G$
is defined as the states in which the master server is up and there is at least one copy of each chunk available. We also consider the occurrence of a severe hardware disaster. The model instances are denoted as ``GFS-$N$(-hw-dis)''.

All model files are available from the repository of IMCA\footnote{\url{http://fmt.cs.utwente.nl/gitweb/imca.git}}, an analyser for MA and IMCs~\cite{GuckHKN12,GuckHHKT13}. Each benchmark instance contains probabilistic states as well as Markovian states, making both kinds of abstraction necessary.

\begin{figure}
  \centering
   \includegraphics[page=5,width=12cm]{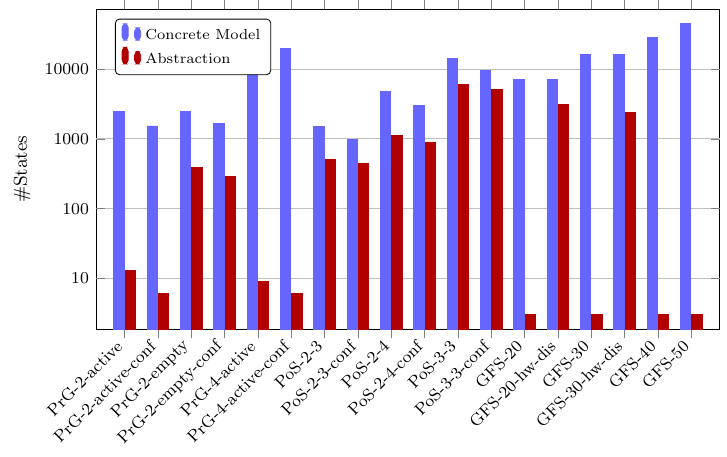}
  \caption{Comparison between the number of states of the concrete and the final abstract model.}
  \label{fig:NumberOfStates}
\end{figure}

Table~\ref{tab:ResultTable} compares our experimental results of the value iteration for the concrete system and with our abstraction refinement framework. We computed the maximum reachability probability $p_\mathit{max}$ for different time bounds. We used precision $\epsilon = 0.01$ for the value iteration as well as for the abstraction refinement.

The first column contains the name of the considered model. The blocks titled ``Concrete Model'' and ``Abstraction'' present the results for the concrete model and for the final result of the abstraction refinement, respectively. The first two columns denote the name of the instance and the applied time bound $\tb$. The third and sixth columns (``\#{}states'') contain the number of states of the concrete model and the final abstraction. Due to the fact that solving a discretised system is rather expensive~\cite{ZhangN10}, the benchmark instances are relatively small.

 Column ``$p$'' denotes the computed maximum probability for the concrete system, whereas ``$\lb$'' and ``$\ub$'' denote the computed lower and upper bounds for the abstraction. Column ``\#{}iter.'' contains the number of iterations of the refinement loop. Column ``time'' states the computation time needed for the analysis of the concrete model, whereas columns ``ref. time'' and ``val. time'' contain the time spent on computing the abstraction refinement and the value iteration. All time measurements are given in the format ``minutes:seconds''. The total computation time needed by the prototype is the sum of the time needed for abstraction refinement and the time needed for the value iteration. As can be seen, the time needed for the abstraction refinement is negligible for the most part.

Computations which took longer than five hours were aborted and are marked with ``(TO)''. All experiments were done on a Dual Core AMD Opteron processor with 2.4 GHz per core and 64 GB of memory. Each computation needed less than 4 GB memory, we therefore do not present measurements of the memory consumption.

For most instances of PrG and GFS the abstraction refinement needs less computation time than the value iteration for the concrete model. For most instances of PoS both approaches need about the same time. For some instances, \eg PoS-3-3, the abstraction refinement is slower than the value iteration. For all case studies we were able to achieve a significant compaction of the state space. The latter is also illustrated in Figure~\ref{fig:NumberOfStates}, which uses a logarithmic scale. If we increase $\epsilon$ and thereby lower the precision, less time is needed for the computation and further compaction is achieved.

\begin{figure}
   \centering
  \includegraphics[page=1,height=0.5cm]{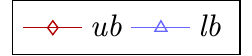}

   \subfigure[]{
      \label{fig:lb_ub_grid}
      \includegraphics[page=2,height=4.4cm]{images/all_plots_smaller_version}
   }
   \subfigure[]{
      \label{fig:lb_ub_polling}
      \includegraphics[page=3,height=4.4cm]{images/all_plots_smaller_version}
   }
   \subfigure[]{
      \label{fig:lb_ub_google}
      \includegraphics[page=4,height=4.4cm]{images/all_plots_smaller_version}
   }
   \caption{Development of $\lb$ and $\ub$.}
   \label{fig:lb_ub_comparision}
\end{figure}

Fig~\ref{fig:lb_ub_comparision} shows the development of the probability bounds $\lb$ and $\ub$ during the abstraction refinement loop for selected instances. The fluctuations which can be seen in the curves for PoS-2-3-conf are due to the increase of accuracy over time.
%%% Local Variables: 
%%% mode: latex
%%% TeX-master: "main"
%%% End: 

% $Id: 05-experiments.tex  braitlin $

\section{Conclusion}
\label{sec:Conclusion}
In this paper we have presented our menu-based game abstraction of MA, which is a combination of successful techniques for the abstraction of PA~\cite{KattenbeltKNP10, Wachter2011}. We also have shown how to analyse the quality of the abstraction for bounded reachability objectives. Should the abstraction turn out to be too coarse, we may refine it using a scheduler-based refinement method which we optimised with a number of additional techniques. Our experiments give promising results and we can report on a significant reduction of the number of states.

As future work we plan to implement a pure game-based abstraction for MA and to compare it to the results of our combined approach. We are also working on the analysis of additional types of properties, \eg expected time of reachability and long-run average. Furthermore, we are going to explore the possibilities of alternative refinement techniques.
%%% Local Variables: 
%%% mode: latex
%%% TeX-master: "main"
%%% End: 

\bibliographystyle{eptcs}
\bibliography{literature}

\begin{thebibliography}{10}
\providecommand{\bibitemdeclare}[2]{}
\providecommand{\surnamestart}{}
\providecommand{\surnameend}{}
\providecommand{\urlprefix}{Available at }
\providecommand{\url}[1]{\texttt{#1}}
\providecommand{\href}[2]{\texttt{#2}}
\providecommand{\urlalt}[2]{\href{#1}{#2}}
\providecommand{\doi}[1]{doi:\urlalt{http://dx.doi.org/#1}{#1}}
\providecommand{\bibinfo}[2]{#2}

\bibitemdeclare{article}{BaierHHK03}
\bibitem{BaierHHK03}
\bibinfo{author}{Christel \surnamestart Baier\surnameend},
  \bibinfo{author}{Boudewijn~R. \surnamestart Haverkort\surnameend},
  \bibinfo{author}{Holger \surnamestart Hermanns\surnameend} \&
  \bibinfo{author}{Joost-Pieter \surnamestart Katoen\surnameend}
  (\bibinfo{year}{2003}): \emph{\bibinfo{title}{Model-Checking Algorithms for
  Continuous-Time Markov Chains}}.
\newblock {\sl \bibinfo{journal}{IEEE Trans.\ on Software Engineering}}
  \bibinfo{volume}{29}(\bibinfo{number}{6}), pp. \bibinfo{pages}{524--541},
  \doi{10.1109/TSE.2003.1205180}.

\bibitemdeclare{article}{BoudaliCS10}
\bibitem{BoudaliCS10}
\bibinfo{author}{Hichem \surnamestart Boudali\surnameend},
  \bibinfo{author}{Pepijn \surnamestart Crouzen\surnameend} \&
  \bibinfo{author}{Mari{\"e}lle \surnamestart Stoelinga\surnameend}
  (\bibinfo{year}{2010}): \emph{\bibinfo{title}{A Rigorous, Compositional, and
  Extensible Framework for Dynamic Fault Tree Analysis}}.
\newblock {\sl \bibinfo{journal}{IEEE Trans. Dependable Sec. Comput.}}
  \bibinfo{volume}{7}(\bibinfo{number}{2}), pp. \bibinfo{pages}{128--143},
  \doi{10.1109/TDSC.2009.45}.

\bibitemdeclare{article}{BozzanoCKNNR11}
\bibitem{BozzanoCKNNR11}
\bibinfo{author}{Marco \surnamestart Bozzano\surnameend},
  \bibinfo{author}{Alessandro \surnamestart Cimatti\surnameend},
  \bibinfo{author}{Joost-Pieter \surnamestart Katoen\surnameend},
  \bibinfo{author}{Viet~Yen \surnamestart Nguyen\surnameend},
  \bibinfo{author}{Thomas \surnamestart Noll\surnameend} \&
  \bibinfo{author}{Marco \surnamestart Roveri\surnameend}
  (\bibinfo{year}{2011}): \emph{\bibinfo{title}{Safety, Dependability and
  Performance Analysis of Extended AADL Models}}.
\newblock {\sl \bibinfo{journal}{Computer Journal}}
  \bibinfo{volume}{54}(\bibinfo{number}{5}), pp. \bibinfo{pages}{754--775},
  \doi{10.1093/comjnl/bxq024}.

\bibitemdeclare{inproceedings}{CosteHLS09}
\bibitem{CosteHLS09}
\bibinfo{author}{Nicolas \surnamestart Coste\surnameend},
  \bibinfo{author}{Holger \surnamestart Hermanns\surnameend},
  \bibinfo{author}{Etienne \surnamestart Lantreibecq\surnameend} \&
  \bibinfo{author}{Wendelin \surnamestart Serwe\surnameend}
  (\bibinfo{year}{2009}): \emph{\bibinfo{title}{Towards Performance Prediction
  of Compositional Models in Industrial GALS Designs}}.
\newblock In: {\sl \bibinfo{booktitle}{Proc.\ of CAV}}, {\sl
  \bibinfo{series}{LNCS}} \bibinfo{volume}{5643},
  \bibinfo{publisher}{Springer}, pp. \bibinfo{pages}{204--218},
  \doi{10.1007/978-3-642-02658-4\_18}.

\bibitemdeclare{inproceedings}{DArgenioJJL02}
\bibitem{DArgenioJJL02}
\bibinfo{author}{Pedro~R. \surnamestart D'Argenio\surnameend},
  \bibinfo{author}{Bertrand \surnamestart Jeannet\surnameend},
  \bibinfo{author}{Henrik~E.\ \surnamestart Jensen\surnameend} \&
  \bibinfo{author}{Kim~G.\ \surnamestart Larsen\surnameend}
  (\bibinfo{year}{2002}): \emph{\bibinfo{title}{Reduction and Refinement
  Strategies for Probabilistic Analysis}}.
\newblock In: {\sl \bibinfo{booktitle}{Proc.\ of PAPM-PROBMIV}}, pp.
  \bibinfo{pages}{57--76}, \doi{10.1007/3-540-45605-8\_5}.

\bibitemdeclare{article}{DengH13}
\bibitem{DengH13}
\bibinfo{author}{Yuxin \surnamestart Deng\surnameend} \&
  \bibinfo{author}{Matthew \surnamestart Hennessy\surnameend}
  (\bibinfo{year}{2013}): \emph{\bibinfo{title}{On the semantics of {Markov}
  automata}}.
\newblock {\sl \bibinfo{journal}{Information and Computation}}
  \bibinfo{volume}{222}, pp. \bibinfo{pages}{139--168},
  \doi{10.1016/j.ic.2012.10.010}.

\bibitemdeclare{inproceedings}{DesharnaisJGP02}
\bibitem{DesharnaisJGP02}
\bibinfo{author}{Josee \surnamestart Desharnais\surnameend},
  \bibinfo{author}{Radha \surnamestart Jagadeesan\surnameend},
  \bibinfo{author}{Vineet \surnamestart Gupta\surnameend} \&
  \bibinfo{author}{Prakash \surnamestart Panangaden\surnameend}
  (\bibinfo{year}{2002}): \emph{\bibinfo{title}{The Metric Analogue of Weak
  Bisimulation for Probabilistic Processes}}.
\newblock In: {\sl \bibinfo{booktitle}{Proc.\ of LICS}},
  \bibinfo{publisher}{IEEE CS}, pp. \bibinfo{pages}{413--422},
  \doi{10.1109/LICS.2002.1029849}.

\bibitemdeclare{inproceedings}{EisentrautHKZ13}
\bibitem{EisentrautHKZ13}
\bibinfo{author}{Christian \surnamestart Eisentraut\surnameend},
  \bibinfo{author}{Holger \surnamestart Hermanns\surnameend},
  \bibinfo{author}{Joost-Pieter \surnamestart Katoen\surnameend} \&
  \bibinfo{author}{Lijun \surnamestart Zhang\surnameend}
  (\bibinfo{year}{2013}): \emph{\bibinfo{title}{A Semantics for Every {GSPN}}}.
\newblock In: {\sl \bibinfo{booktitle}{Proc. of Petri Nets}}, {\sl
  \bibinfo{series}{LNCS}} \bibinfo{volume}{7927},
  \bibinfo{publisher}{Springer}, pp. \bibinfo{pages}{90--109},
  \doi{10.1007/978-3-642-38697-8\_6}.

\bibitemdeclare{inproceedings}{EisentrautHZ10:concur}
\bibitem{EisentrautHZ10:concur}
\bibinfo{author}{Christian \surnamestart Eisentraut\surnameend},
  \bibinfo{author}{Holger \surnamestart Hermanns\surnameend} \&
  \bibinfo{author}{Lijun \surnamestart Zhang\surnameend}
  (\bibinfo{year}{2010}): \emph{\bibinfo{title}{Concurrency and Composition in
  a Stochastic World}}.
\newblock In: {\sl \bibinfo{booktitle}{Proc.\ of CONCUR}}, {\sl
  \bibinfo{series}{LNCS}} \bibinfo{volume}{6269},
  \bibinfo{publisher}{Springer}, pp. \bibinfo{pages}{21--39},
  \doi{10.1007/978-3-642-15375-4\_3}.

\bibitemdeclare{inproceedings}{EisentrautHZ10:lics}
\bibitem{EisentrautHZ10:lics}
\bibinfo{author}{Christian \surnamestart Eisentraut\surnameend},
  \bibinfo{author}{Holger \surnamestart Hermanns\surnameend} \&
  \bibinfo{author}{Lijun \surnamestart Zhang\surnameend}
  (\bibinfo{year}{2010}): \emph{\bibinfo{title}{On Probabilistic Automata in
  Continuous Time}}.
\newblock In: {\sl \bibinfo{booktitle}{Proc.\ of LICS}},
  \bibinfo{publisher}{IEEE CS}, pp. \bibinfo{pages}{342--351},
  \doi{10.1109/LICS.2010.41}.

\bibitemdeclare{inproceedings}{GhemawatGL03}
\bibitem{GhemawatGL03}
\bibinfo{author}{Sanjay \surnamestart Ghemawat\surnameend},
  \bibinfo{author}{Howard \surnamestart Gobioff\surnameend} \&
  \bibinfo{author}{Shun-Tak \surnamestart Leung\surnameend}
  (\bibinfo{year}{2003}): \emph{\bibinfo{title}{The Google file system}}.
\newblock In: {\sl \bibinfo{booktitle}{Proc. of the ACM Symp. on Operating
  Systems Principles (SOSP)}}, \bibinfo{publisher}{ACM Press}, pp.
  \bibinfo{pages}{29--43}, \doi{10.1145/945445.945450}.

\bibitemdeclare{mastersthesis}{Guck12}
\bibitem{Guck12}
\bibinfo{author}{D.~\surnamestart {Guck}\surnameend} (\bibinfo{year}{2012}):
  \emph{\bibinfo{title}{Quantitative Analysis of Markov Automata}}.
\newblock Master's thesis, \bibinfo{school}{RWTH Aachen University}.

\bibitemdeclare{inproceedings}{GuckHKN12}
\bibitem{GuckHKN12}
\bibinfo{author}{Dennis \surnamestart Guck\surnameend},
  \bibinfo{author}{Tingting \surnamestart Han\surnameend},
  \bibinfo{author}{Joost-Pieter \surnamestart Katoen\surnameend} \&
  \bibinfo{author}{Martin~R. \surnamestart Neuh{\"a}u{\ss}er\surnameend}
  (\bibinfo{year}{2012}): \emph{\bibinfo{title}{Quantitative Timed Analysis of
  Interactive Markov Chains}}.
\newblock In: {\sl \bibinfo{booktitle}{Proc.\ of NFM}}, {\sl
  \bibinfo{series}{LNCS}} \bibinfo{volume}{7226},
  \bibinfo{publisher}{Springer}, pp. \bibinfo{pages}{8--23},
  \doi{10.1007/978-3-642-28891-3\_4}.

\bibitemdeclare{inproceedings}{GuckHHKT13}
\bibitem{GuckHHKT13}
\bibinfo{author}{Dennis \surnamestart Guck\surnameend}, \bibinfo{author}{Hassan
  \surnamestart Hatefi\surnameend}, \bibinfo{author}{Holger \surnamestart
  Hermanns\surnameend}, \bibinfo{author}{Joost-Pieter \surnamestart
  Katoen\surnameend} \& \bibinfo{author}{Mark \surnamestart Timmer\surnameend}
  (\bibinfo{year}{2013}): \emph{\bibinfo{title}{Modelling, Reduction and
  Analysis of {Markov} Automata}}.
\newblock In: {\sl \bibinfo{booktitle}{Proc.\ of QEST}}, {\sl
  \bibinfo{series}{LNCS}} \bibinfo{volume}{8054},
  \bibinfo{publisher}{Springer}, pp. \bibinfo{pages}{55--71},
  \doi{10.1007/978-3-642-40196-1\_5}.

\bibitemdeclare{article}{HatefiH12}
\bibitem{HatefiH12}
\bibinfo{author}{Hassan \surnamestart Hatefi\surnameend} \&
  \bibinfo{author}{Holger \surnamestart Hermanns\surnameend}
  (\bibinfo{year}{2012}): \emph{\bibinfo{title}{Model Checking Algorithms for
  {Markov} Automata}}.
\newblock {\sl \bibinfo{journal}{ECEASST}} \bibinfo{volume}{53},
  \doi{10.1007/978-3-642-40213-5\_16}.

\bibitemdeclare{inproceedings}{HaverkortKRRS10}
\bibitem{HaverkortKRRS10}
\bibinfo{author}{Boudewijn~R. \surnamestart Haverkort\surnameend},
  \bibinfo{author}{Matthias \surnamestart Kuntz\surnameend},
  \bibinfo{author}{Anne \surnamestart Remke\surnameend},
  \bibinfo{author}{S.~\surnamestart Roolvink\surnameend} \&
  \bibinfo{author}{Mari{\"e}lle \surnamestart Stoelinga\surnameend}
  (\bibinfo{year}{2010}): \emph{\bibinfo{title}{Evaluating repair strategies
  for a water-treatment facility using Arcade}}.
\newblock In: {\sl \bibinfo{booktitle}{Proc. of DSN}},
  \bibinfo{publisher}{IEEE}, pp. \bibinfo{pages}{419--424},
  \doi{10.1109/DSN.2010.5544290}.

\bibitemdeclare{book}{Hermanns02}
\bibitem{Hermanns02}
\bibinfo{author}{Holger \surnamestart Hermanns\surnameend}
  (\bibinfo{year}{2002}): \emph{\bibinfo{title}{Interactive {Markov} Chains --
  The Quest for Quantified Quality}}.
\newblock {\sl \bibinfo{series}{LNCS}} \bibinfo{volume}{2428},
  \bibinfo{publisher}{Springer}, \doi{10.1007/3-540-45804-2}.

\bibitemdeclare{article}{KattenbeltKNP10}
\bibitem{KattenbeltKNP10}
\bibinfo{author}{Mark \surnamestart Kattenbelt\surnameend},
  \bibinfo{author}{Marta~Z. \surnamestart Kwiatkowska\surnameend},
  \bibinfo{author}{Gethin \surnamestart Norman\surnameend} \&
  \bibinfo{author}{David \surnamestart Parker\surnameend}
  (\bibinfo{year}{2010}): \emph{\bibinfo{title}{A game-based
  abstraction-refinement framework for {Markov} decision processes}}.
\newblock {\sl \bibinfo{journal}{Formal Methods in System Design}}
  \bibinfo{volume}{36}(\bibinfo{number}{3}), pp. \bibinfo{pages}{246--280},
  \doi{10.1007/s10703-010-0097-6}.

\bibitemdeclare{article}{MarsanCB84}
\bibitem{MarsanCB84}
\bibinfo{author}{Marco~Ajmone \surnamestart Marsan\surnameend},
  \bibinfo{author}{Gianni \surnamestart Conte\surnameend} \&
  \bibinfo{author}{Gianfranco \surnamestart Balbo\surnameend}
  (\bibinfo{year}{1984}): \emph{\bibinfo{title}{A Class of Generalized
  Stochastic Petri Nets for the Performance Evaluation of Multiprocessor
  Systems}}.
\newblock {\sl \bibinfo{journal}{ACM Trans. Comput. Syst.}}
  \bibinfo{volume}{2}(\bibinfo{number}{2}), pp. \bibinfo{pages}{93--122},
  \doi{10.1145/190.191}.

\bibitemdeclare{inproceedings}{MeyerMS85}
\bibitem{MeyerMS85}
\bibinfo{author}{John~F. \surnamestart Meyer\surnameend}, \bibinfo{author}{Ali
  \surnamestart Movaghar\surnameend} \& \bibinfo{author}{William~H.
  \surnamestart Sanders\surnameend} (\bibinfo{year}{1985}):
  \emph{\bibinfo{title}{Stochastic Activity Networks: Structure, Behavior, and
  Application}}.
\newblock In: {\sl \bibinfo{booktitle}{Proc. of PNPM}},
  \bibinfo{publisher}{IEEE CS}, pp. \bibinfo{pages}{106--115}.

\bibitemdeclare{phdthesis}{Neuhausser10}
\bibitem{Neuhausser10}
\bibinfo{author}{Martin~R. \surnamestart Neuh{\"a}u{\ss}er\surnameend}
  (\bibinfo{year}{2010}): \emph{\bibinfo{title}{Model checking nondeterministic
  and randomly timed systems}}.
\newblock Ph.D. thesis, \bibinfo{school}{RWTH Aachen University and University
  of Twente}.

\bibitemdeclare{phdthesis}{Segala95}
\bibitem{Segala95}
\bibinfo{author}{Roberto \surnamestart Segala\surnameend}
  (\bibinfo{year}{1995}): \emph{\bibinfo{title}{Modeling and Verification of
  Randomized Distributed Real-Time Systems}}.
\newblock Ph.D. thesis, \bibinfo{school}{MIT}.

\bibitemdeclare{article}{Shapley1953}
\bibitem{Shapley1953}
\bibinfo{author}{Lloyd~S \surnamestart Shapley\surnameend}
  (\bibinfo{year}{1953}): \emph{\bibinfo{title}{Stochastic games}}.
\newblock {\sl \bibinfo{journal}{Proceedings of the National Academy of
  Sciences of the United States of America}}
  \bibinfo{volume}{39}(\bibinfo{number}{10}), p. \bibinfo{pages}{1095},
  \doi{10.1073/pnas.39.10.1095}.

\bibitemdeclare{inproceedings}{TimmerKPS12}
\bibitem{TimmerKPS12}
\bibinfo{author}{Mark \surnamestart Timmer\surnameend},
  \bibinfo{author}{Joost-Pieter \surnamestart Katoen\surnameend},
  \bibinfo{author}{Jaco \surnamestart van~de Pol\surnameend} \&
  \bibinfo{author}{Mari{\"e}lle \surnamestart Stoelinga\surnameend}
  (\bibinfo{year}{2012}): \emph{\bibinfo{title}{Efficient Modelling and
  Generation of Markov Automata}}.
\newblock In: {\sl \bibinfo{booktitle}{Proc.\ of CONCUR}}, {\sl
  \bibinfo{series}{LNCS}} \bibinfo{volume}{7454},
  \bibinfo{publisher}{Springer}, pp. \bibinfo{pages}{364--379},
  \doi{10.1007/978-3-642-32940-1\_26}.

\bibitemdeclare{inproceedings}{TimmerPS13}
\bibitem{TimmerPS13}
\bibinfo{author}{Mark \surnamestart Timmer\surnameend}, \bibinfo{author}{Jaco
  \surnamestart van~de Pol\surnameend} \& \bibinfo{author}{Mari{\"e}lle
  \surnamestart Stoelinga\surnameend} (\bibinfo{year}{2013}):
  \emph{\bibinfo{title}{Confluence Reduction for {Markov} Automata}}.
\newblock In: {\sl \bibinfo{booktitle}{Proc.\ of FORMATS}}, {\sl
  \bibinfo{series}{LNCS}} \bibinfo{volume}{8053},
  \bibinfo{publisher}{Springer}, pp. \bibinfo{pages}{243--257},
  \doi{10.1007/978-3-642-40229-6\_17}.

\bibitemdeclare{phdthesis}{Wachter2011}
\bibitem{Wachter2011}
\bibinfo{author}{Bj{\"o}rn \surnamestart Wachter\surnameend}
  (\bibinfo{year}{2011}): \emph{\bibinfo{title}{Refined probabilistic
  abstraction}}.
\newblock Ph.D. thesis, \bibinfo{school}{Saarland University}.

\bibitemdeclare{inproceedings}{WachterZ10}
\bibitem{WachterZ10}
\bibinfo{author}{Bj\"{o}rn \surnamestart Wachter\surnameend} \&
  \bibinfo{author}{Lijun \surnamestart Zhang\surnameend}
  (\bibinfo{year}{2010}): \emph{\bibinfo{title}{Best Probabilistic
  Transformers}}.
\newblock In: {\sl \bibinfo{booktitle}{Proc.\ of VMCAI}}, {\sl
  \bibinfo{series}{LNCS}} \bibinfo{volume}{5944},
  \bibinfo{publisher}{Springer}, pp. \bibinfo{pages}{362--379},
  \doi{10.1007/978-3-642-11319-2\_26}.

\bibitemdeclare{inproceedings}{ZhangN10}
\bibitem{ZhangN10}
\bibinfo{author}{Lijun \surnamestart Zhang\surnameend} \&
  \bibinfo{author}{Martin~R. \surnamestart Neuh{\"a}u{\ss}er\surnameend}
  (\bibinfo{year}{2010}): \emph{\bibinfo{title}{Model Checking Interactive
  Markov Chains}}.
\newblock In: {\sl \bibinfo{booktitle}{Proc.\ of TACAS}}, {\sl
  \bibinfo{series}{LNCS}} \bibinfo{volume}{6015},
  \bibinfo{publisher}{Springer}, pp. \bibinfo{pages}{53--68},
  \doi{10.1007/978-3-642-12002-2\_5}.

\end{thebibliography}

\end{document}